\documentclass[conference]{IEEEtran}
\IEEEoverridecommandlockouts
\usepackage{cite}
\usepackage{amsmath,amssymb,amsfonts}
\usepackage{algorithmic}
\usepackage{graphicx}
\usepackage{textcomp}
\usepackage{xcolor}
\def\BibTeX{{\rm B\kern-.05em{\sc i\kern-.025em b}\kern-.08em
    T\kern-.1667em\lower.7ex\hbox{E}\kern-.125emX}}

\usepackage{comment}
\usepackage{soul}
\usepackage{makecell}
\usepackage{url}

\title{Design and benchmarking of a two degree of freedom tendon driver unit for cable-driven wearable technologies}
\author{Adrian Esser$^{1,*}$, Chiara Basla$^{1}$, Peter Wolf$^{1}$, Robert Riener$^{1,2}$

\thanks{This work was supported as a part of NCCR Robotics, a National Centre of Competence in Research, funded by the Swiss National Science Foundation (grant number 51NF40\_185543).}
\hspace*{-1cm} \thanks{$^{1}$ Sensory-Motor Systems Lab, Institute of Robotics and Intelligent Systems (IRIS), ETH Zurich, Switzerland}%
\thanks{$^{2}$ Spinal Cord Injury Center, Balgrist University Hospital, Medical Faculty, University of Zürich, Switzerland}%%
\thanks{$^{*}$ Corresponding author: adrian.esser@hest.ethz.ch}%%
}

\begin{document}

\maketitle

\begin{abstract}
Exosuits have recently been developed as alternatives to rigid exoskeletons and are increasingly adopted for both upper and lower limb therapy and assistance in clinical and home environments. Many cable-driven exosuits have been developed but little has been published on their electromechanical designs and performance. Therefore, this paper presents a comprehensive design and performance analysis of a two degree of freedom tendon driver unit (TDU) for cable-driven wearable exosuits. Detailed methodologies are presented to benchmark the functionality of the TDU. A static torque output test compares the commanded and measured torques. A velocity control test evaluates the attenuation and phase shift across velocities. A noise test evaluates how loud the TDU is for the wearer under different speeds. A thermal stress test captures the cooling performance of the TDU to ensure safe operation at higher loads. Finally, a battery endurance test evaluates the runtime of the TDU under various loading conditions to inform the usable time. To demonstrate these tests, a modular TDU system for cable-driven applications is introduced, which allows components such as motors, pulleys, and sensors to be adapted based on the requirements of the intended application. By sharing detailed methodologies and performance results, this study aims to provide a TDU design that may be leveraged by others and resources for researchers and engineers to better document the capabilities of their TDU designs.
\end{abstract}


\section{Introduction}
In the past decade, exosuits have emerged as a technology to support individuals with partial motor functions, enhancing their ability to participate in daily activities. Unlike traditional rigid exoskeletons, exosuits offer a lighter, more flexible, and affordable solution that prioritizes user comfort and accessibility for everyday use. These qualities make exosuits versatile tools with potential for broad adoption across multiple applications. Based on actuation methodologies, exosuits are classified into cable-driven, pneumatic, twisted strings, and shape-memory alloy systems \cite{xiloyannis}. Recent advances in electric motor control and the flexible routing of cables have expanded cable-driven exosuits, which are currently the most widely adopted form \cite{xiloyannis}.

\begin{figure}
    \centering
    \includegraphics[trim={8.5cm 0cm 7.5cm 0cm},clip,width=0.95\linewidth]{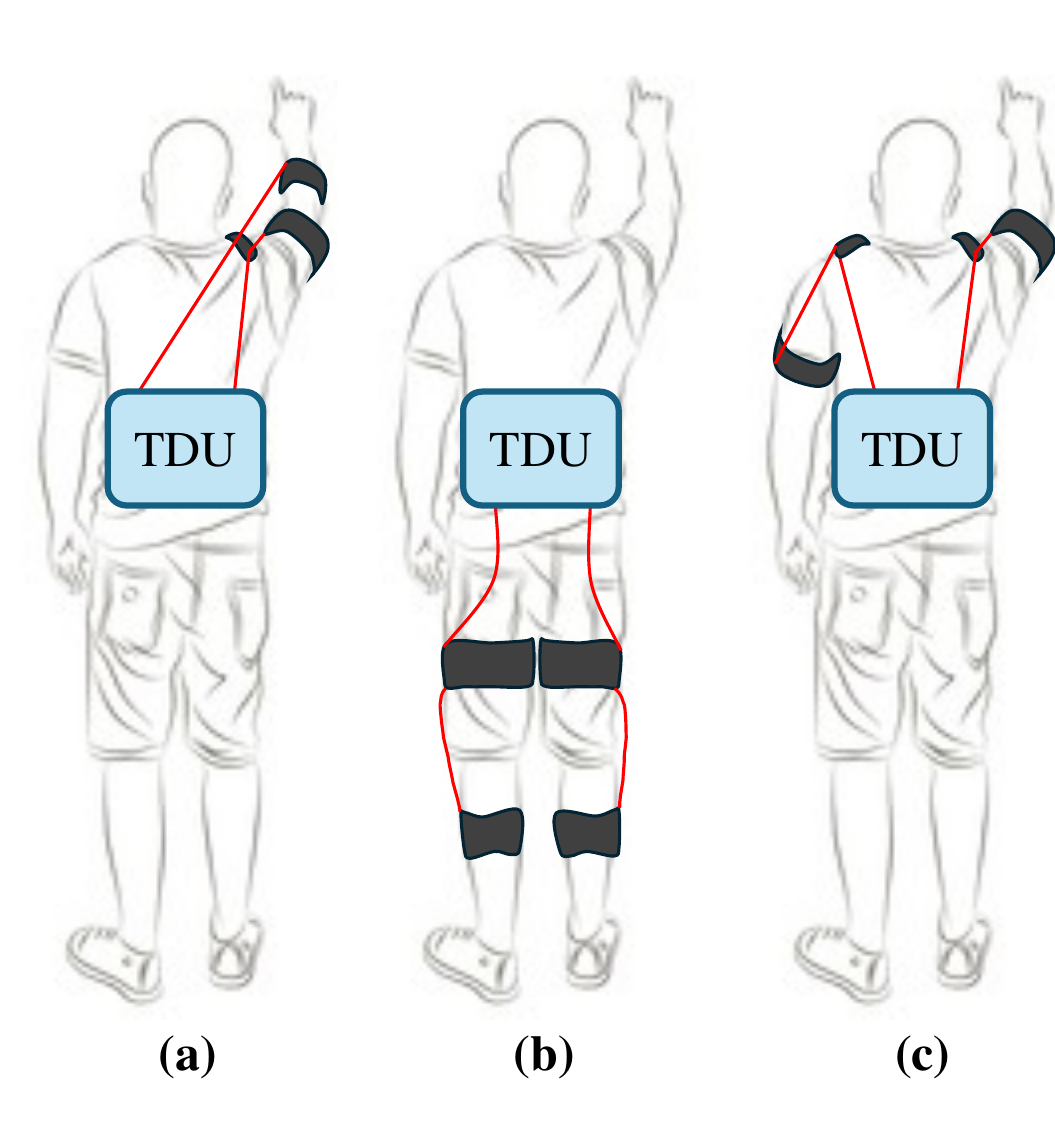}
    \caption{Three conceptual designs demonstrating examples of how a two DOF TDU may support multiple joints: \textbf{(a)} shoulder and elbow on one arm, \textbf{(b)} hip and/or knee on the legs, or \textbf{(c)} shoulder elevation on both arms.}
    \label{fig:conceptual_designs}
\end{figure}

Exosuits have been developed for both lower and upper limb assistance. Lower limb devices often actuate single joints, like the ankle \cite{awad,bae} or hip \cite{asbeck,panizzolo,kim}, and can also integrate multi-joint actuation for comprehensive support, such as hip-ankle \cite{asbeck2} or hip-knee systems \cite{schmidt,park,chen}. For upper limbs, exosuits typically have one or two actuated joints with a range of configurations to support various arm movements effectively. Examples include elbow flexion \cite{nycz}, shoulder-elbow flexion \cite{pont,samperescudero}, wrist support \cite{choi}, and shoulder elevation \cite{georgarakisNature,bardiesser}. Each of these examples has one or two degrees of freedom (DOF), referring to the number of actuators embedded in the exosuit. Each actuator can affect multiple joints and facilitate multidirectional movement based on tendon routing along the limb (Fig. \ref{fig:conceptual_designs}).

 %Strategically positioning the TDU near the user’s center of mass minimizes load on distal joints, enhancing both comfort and mobility. Flexible cable routing and attachments allow natural movement with minimal restrictions.

All portable cable-driven exosuits include an actuation system, referred to as tendon driver unit (TDU). This commonly houses electronics, motors, batteries, and one or more cables routed to flexible joint attachments. Despite notable progress, documentation and transparent sharing of exosuit TDU design, performance metrics, and benchmarks remain limited. This lack of detail impedes researchers and engineers from replicating existing systems, refining devices, and building on shared knowledge within the field.

To address this gap, we outline benchmarking tests to evaluate performance. In addition to standard performance metrics, such as torque and velocity, we recommend further tests to assess noise, battery life, and thermal management, all critical considerations for portable exosuit applications suited for clinic or home use. Factors such as torque and velocity performance are important to evaluate because they fundamentally determine the range of anthropometrics that can be supported by the device (i.e., how heavy or tall the user is) and what types of movements may be performed (i.e. how fast one can lift the arm or walk). Factors such as noise, battery life, and thermal management are included as well as they are often overlooked, but also are necessary for device usability and acceptance. By sharing these methodologies, we aim to foster performance transparency and comparability within the exosuit field, facilitating improved design choices, accelerated advancements, and enriched technical documentation through accessible research and open-source repositories.\footnote{All design files and scripts for testing and data processing are available publicly: \url{https://bitbucket.org/AdrianJEsser/icorr_2025_tdu/src/main/}}

Alongside the methodologies, we present a lightweight, adaptable TDU design with two DOF. This TDU supports multiple configurations, making it suitable for both upper and lower limb exosuits. It can be customized to accommodate different applications and user needs through modular features that enable the swapping of components to adjust battery capacity, voltage, and output torque, along with a flexible programming interface to support diverse performance and sensor requirements.

\begin{figure}
    \centering
    \includegraphics[width=0.95\linewidth]{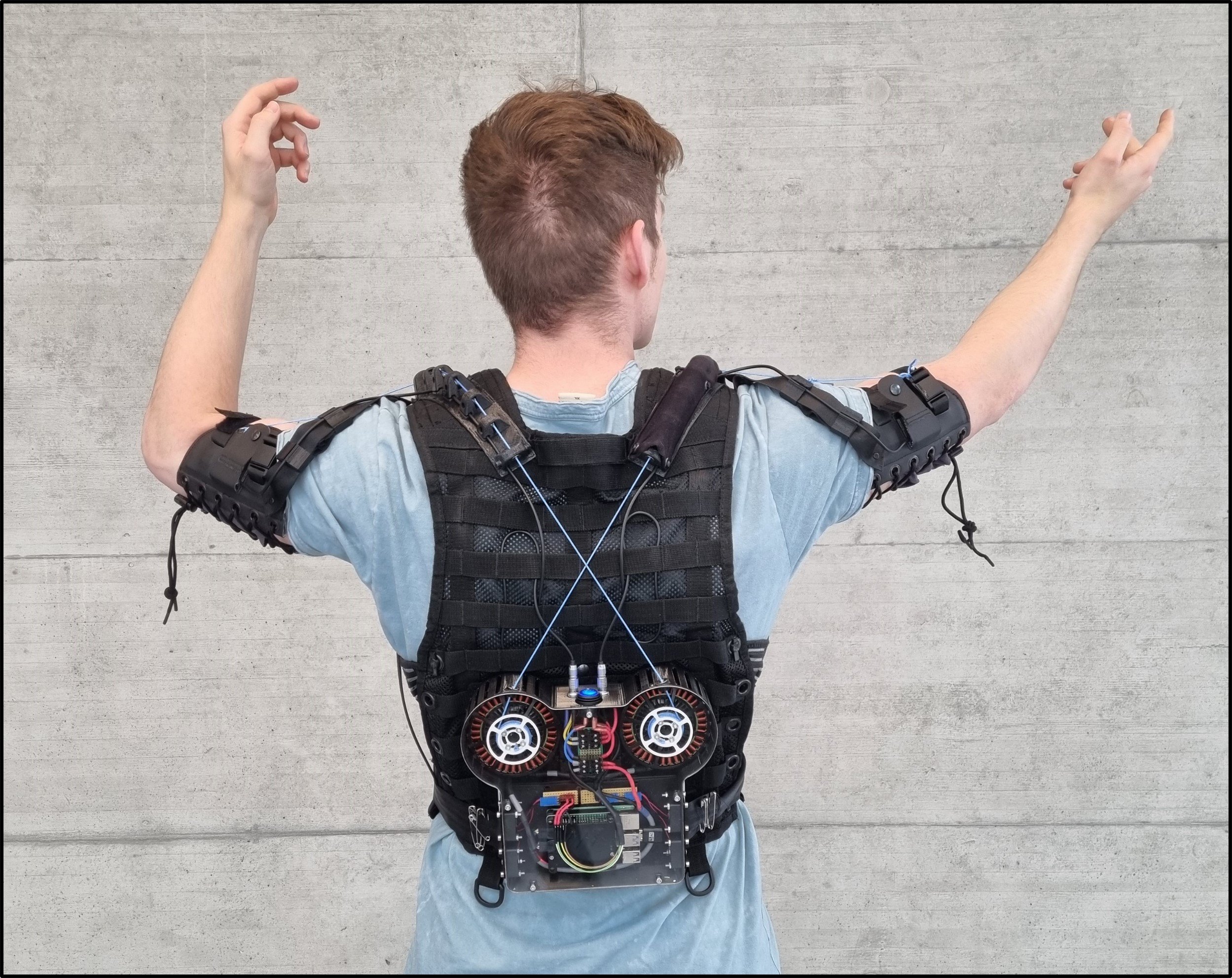}
    \caption{An example of how the TDU can be mounted on the body in real life. In this application, the TDU is used, along with a textile interface, to support the right and left shoulder joints against gravity during shoulder elevation.}
    \label{fig:real_life_example}
\end{figure}

%Some ideas for the main points / motivation behind the submission:

%\begin{itemize}
 %   \item Electromechanical designs and performance capabilities for tendon driver units (also systems and protocols in general) are often poorly documented, hindering replication and sharing of designs and insights.
 %  \item A simple and compact design to be shared with the community that can be refined / modified / integrated based on the needs of the particular application or patient population (stronger / weaker motors, more or less motors, more / less battery, flexible computing program for additional sensors and complex control algorithms)
 %  \item Can be applied to the upper limbs and lower limbs, can be used for bimanual support or unimanual support (and 2 DOF)
 %  \item Suggestion for some benchmarking tests to share capabilities of this particular TDU, but also give researchers ideas for how they may benchmark the performance capabilities of their designs (which can also be reported in technical publications, or released in public repositories)
%\end{itemize}

\section{Tendon Driver Unit Design}

\subsection{Description of Components}
\begin{figure}
    \centering
    \includegraphics[trim={0cm 0cm 0cm 5cm},clip,width=0.95\linewidth]{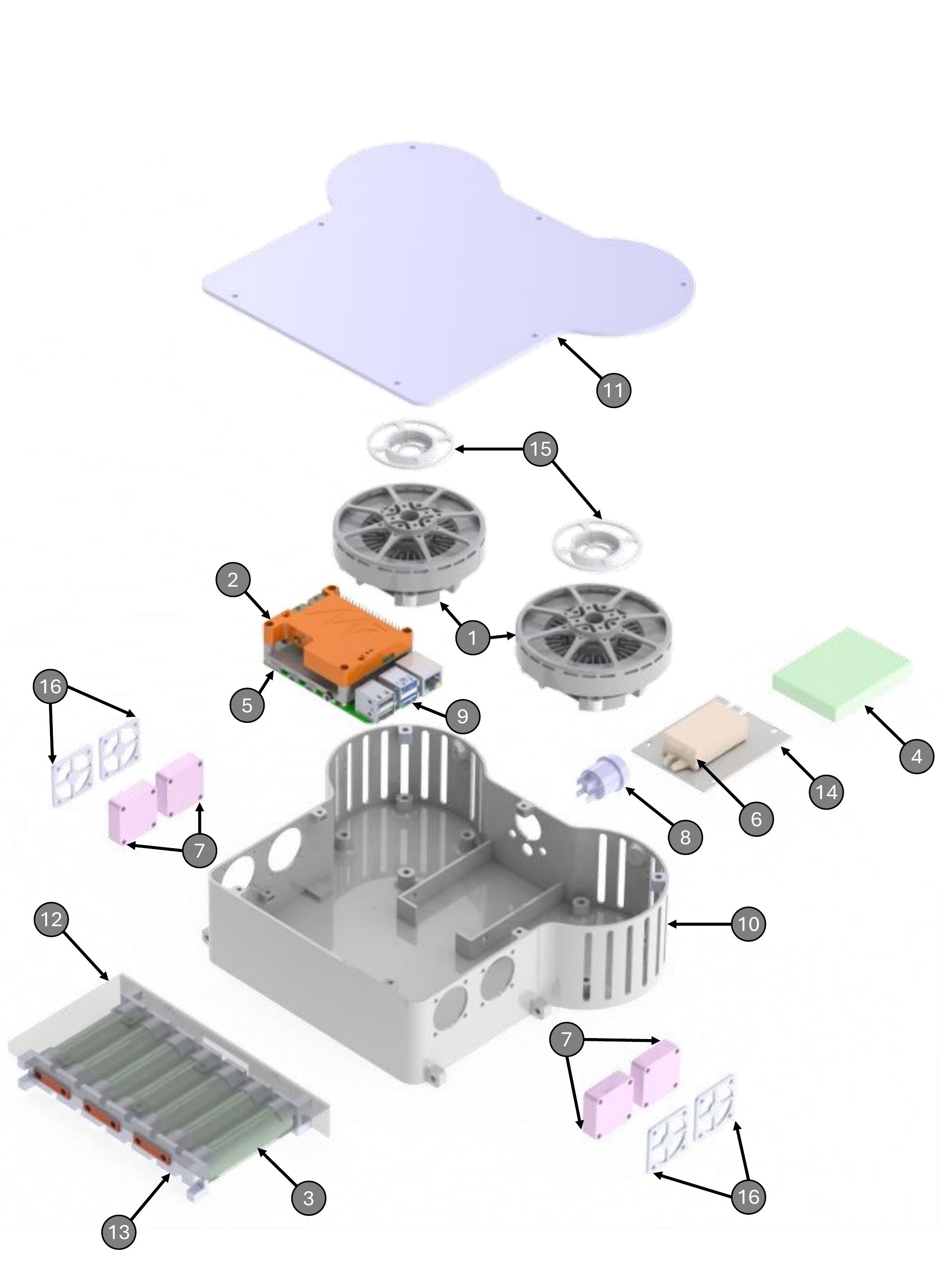}
    \caption{Rendering of an exploded view of the TDU. The numbers correspond to the system components in Table \ref{tab:mass_breakdown}.}
    \label{fig:tdu_exploded}
\end{figure}

The main body of the TDU was 3D printed (Prusa MK4S) out of black polylactic acid (PLA) (Fig. \ref{fig:tdu_exploded} and Tab. \ref{tab:mass_breakdown}). Other custom designed printed parts include the protective covers for the fans, the 30mm cable pulleys, the cover for the battery management system (BMS), and the cover for the battery module. The cover for the TDU was laser cut (Trotec Speedy 400) with 3mm thick transparent acrylic. The 7S1P battery pack (24V. nominal) is custom assembled from LG MJ1 18650 battery cells (3500mAh) and a 24V. 7S BMS (Vruzend). Controlling the TDU is a Raspberry Pi 4B (8Gb RAM) with a RTL881cu USB WiFi network card (PiShop.ch), allowing the TDU to simultaneously connect to a Wifi network and broadcast an independent access point. Power is provided to the microcontroller with a 5V 3A step down buck converter module (PiShop.ch). The regulated 5V module also powers the four PWM HighPi brushless fans (1.4CFM per fan, PiShop.ch), which pull in air from the side of the case, and expel it near the top by the motors. The TDU houses two U8 Lite L KV95 direct drive motors (1.5Nm rated torque, 3Nm peak torque, from MAB Robotics) which are controlled with MD80 brushless motor controller drives (MAB Robotics). Communication to the motors is performed over a controller area network (CAN) protocol with a CANdle hat for the Raspberry Pi (MAB Robotics).

\begin{table}
    \centering
    \caption{Mass breakdown for all major components of the TDU. The numbering in the ID column corresponds to the labels in the exploded view presented in Fig. \ref{fig:tdu_exploded}.}
    \begin{tabular}{|r|l|r|}
        \hline
        ID & Components & Weight (g)\\
        \hline \hline
        & \textbf{Actuation} & - \\
        1 & 2x U8 Lite L KV95 + MD80 & 636 \\
        2 & CANdle Hat & 28 \\
        & Hardware and wiring & 52 \\
        & \textbf{Subtotal Actuation} & \textbf{716} \\
        \hline
        & \textbf{Battery} & - \\
        3 & 7x LG INR18650MJ1 Cells & 343 \\
        4 & 24V 7S BMS & 80  \\
        & Hardware and wiring & 28 \\
        & \textbf{Subtotal Battery} & \textbf{447} \\
        \hline
        & \textbf{Electronics} & - \\
        5 & Raspberry Pi 4B & 46 \\
        6 & 5V converter module & 32 \\
        7 & Cooling fans & 20 \\
        8 & Power button & 5 \\
        9 & USB Wifi card & 3 \\
        & Hardware and wiring & 24 \\
        & \textbf{Subtotal Electronics} & \textbf{134} \\
        \hline
        & \textbf{Printed Parts} & - \\
        10 & Main housing & 237 \\
        11 & Acrylic cover & 118 \\
        12 & Battery cell cover & 42 \\
        13 & Battery cell holders & 26 \\
        14 & BMS cover & 11 \\
        15 & 2x 30mm pulleys & 6 \\
        16 & Fan covers & 5 \\
        & Hardware & 32 \\
        & \textbf{Subtotal Printed Parts} & \textbf{477} \\
        \hline
        & \textbf{Grand Total} & \textbf{1774} \\
        \hline 
    \end{tabular}
    \vspace{0.2cm}
    \label{tab:mass_breakdown}
\end{table}

\begin{figure}
    \centering
    \includegraphics[trim={0cm 0cm 0cm 0cm},clip,width=0.95\linewidth]{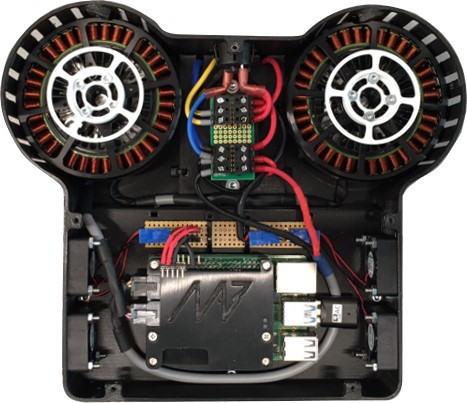}
    \caption{Aerial view of the assembled TDU}
    \label{fig:tdu_top}
\end{figure}

\section{Performance Testing Methodologies}

The protocols for all performance tests are detailed in the subsections below. The left and right motors are referred to as Motor 1 and Motor 2 respectively, when viewing Fig. \ref{fig:tdu_top}.

The open-loop torque controllers and close-loopthe mea velocity controllers used in the following performance tests were not developed by the authors, but used directly from the manufacturers of the motor and controllers, available here: \url{https://mabrobotics.github.io/MD80-x-CANdle-Documentation/intro.html}

\subsection{Open-Loop Static Torque Performance}
This test evaluates the precision and accuracy of the motors when producing static torques. Actuator torque is not directly measured using the MD80 controller, but rather estimated by measuring the motor phase currents. This can be a suitable proxy for the torque of the drive when a low gear ratio is used (i.e. the motor is backdrivable) \cite{lee}. For this test, a pulley with a diameter of 3cm was printed and fixed on the rotor of the TDU motor. The TDU was fixed horizontally, and connected to a digital crane scale (ZW 5.0 force gauge from HKM Messtechnik, 500N nominal, 0.1N step) via a cable (1.5mm Dyneema, extremtextil). A series of increasing fixed torques were commanded (0.25Nm to 2.0Nm in 0.25Nm increments), the system was given 30s to settle, and the load on the scale was recorded. Five repetitions were performed for each condition to ascertain the repeatability of the system. The measured tensile load from the crane scale was converted to the actual output torque of the motor given the pulley diameter of 3cm, and this was compared to the commanded torque of the motor. The protocol was repeated twice, once for each motor in the TDU. 

\subsection{Closed-Loop Velocity Control}
This test evaluates the attenuation and phase shift of the motor when tracking velocity reference signals with varying frequencies and amplitudes. This is an important test to understand if a particular actuator can meet the requirement speeds and frequencies of the intended application. Sinusoidal velocity reference signals were commanded to the drives while in velocity PID control. The signals had amplitudes ranging from 5rad/s up to 30rad/s in 5rad/s increments and 20 frequencies ranging from 0.1Hz to 10Hz in even log-space increments. For each commanded signal 10 full cycles of the motor response velocity were collected. Then the gain and phase shift of the actual motor velocity was computed relative to the reference velocity signal and averaged between both motors. The tests were done without the pulleys attached to the rotors.

\subsection{Thermal Cooling}
This test evaluates the thermal cooling effect of the fans in the TDU. Thermal management is essential for safe continued operation of motors under higher loads and for longer times. The TDU was fixed flat on a table, and both motors were connected to each other outside of the TDU with a re-direct pulley. The left motor was controlled in torque to a constant 1.5Nm (rated torque from the manufacturer), while the right motor was controlled in position executing a sinusoidal profile with an amplitude of $2\pi$rad and a period of 4s. The test was performed dynamically to ensure even heating of all motor windings. The motor stator temperature was recorded with a NTC100K thermistor integrated with MD80 controller. The TDU was turned on with all four fans running (100\% duty cycle), and given 5min for the motor temperature to stabilize with the airflow. Then the system was held dynamically under tension until the first of the two motors hit 80$^\circ$C, then the tension was released from the system and the motors were allowed to cool. Then the fans were disabled, and the test was repeated again without the cooling airflow.  

\subsection{Noise}
This test evaluates the noise level of the TDU in various dynamic conditions. Noise level can impact usability and comfort if too loud by obstructing conversations, and in extreme cases lead to gradual hearing loss. Noise level testing was done in a sound isolated room using a NTI XL2 audio analyzer. The microphone was placed horizontally on a table, with the head of the microphone 50cm away from the top of the TDU, and centered between the two motors. The TDU was lying flat on the table with the acrylic cover facing up. Both the microphone and TDU were set on a thin piece of foam to prevent vibrations from passing through the table. All tests were triggered remotely from outside of the isolated room.

For all tests, the equivalent continuous sound pressure level was recorded over a 20 second window. First, the sound floor of the room was captured with the TDU off. Second, the noise level was captured with the TDU on with fans (100\% duty cycle) but both motors disabled and stationary. Finally, the noise level was recorded for both motors, individually and together, over the set of speeds: 5, 10, 15, 20, 25 and 30 rad/s. 

For all measurements the sound floor of the room with the TDU off ($L_{room}$) was subtracted from the noise level of the TDU in its respective operating condition ($L_{meas}$) using the following equation: 

\begin{equation}
    L_{meas} - L_{room} = 10 \cdot \text{log}_{10}(10^{\frac{L_{meas}}{10}} - 10^{\frac{L_{room}}{10}})
\end{equation}

\subsection{Battery Life}
Evaluating the battery life under different loading conditions can give developers better guidance in specifying motors and battery cells in TDU design. The runtime of the TDU was tested for four different conditions: idle, 2kg, 4kg, and 6kg dynamic loading. The three loading conditions correspond to theoretical average motor torques of 0.29Nm, 0.59Nm, and 0.88Nm respectively, spanning the usable torque range of the motors. For the idle condition, the TDU was left on with the microcontroller and fans running, but the two motors disabled. For the dynamic loading conditions, the TDU raised and lowered the weights according to a sinusoidal position profile with an amplitude of $2\pi$rad and a period of 6s. The TDU was fixed horizontally on a table, and the weights were lifted vertically, using a re-direct pulley mounted on the edge of the table to convert the horizontal pulling force to a vertical force.  For all tests, the TDU was first fully charged (29.1V) and then we measured the time it took to discharge until the BMS hit the low voltage cutoff (17.5V) after which the TDU would shut off to protect the cells from over-discharge. 

\section{Results and Discussion}

All tests were done with the TDU in a horizontal orientation, but as Figure \ref{fig:conceptual_designs} and \ref{fig:real_life_example} suggest, when the TDU is in use it would likely have a vertical mounting on the body (so that the tendons exit either towards the legs or the shoulders). In general, loads were applied to the motors using redirect pulleys so that the loads could be moved against gravity, but the TDU could be mounted and secured on the table in a horizontal orientation (for practicality of performing the test in a controlled environment). It is likely that the re-direct pulleys introduced a small amount of friction to the test setup, so the true performance of the TDU might be slightly better than these test results suggest (across all tests that involved loading the TDU motors). Despite this, the additional friction that would be introduced to the cable transmission system, for example in Figure \ref{fig:real_life_example} where the tendons are routed from the back all the way over the shoulders, should be much greater than the friction introduced by a single re-direct pulley used for these tests. Thus it is anticipated that these results still present an upper bound on the performance of the TDU unit (for this configuration of batteries and motors), and the general methodologies still provide a meaningful method to bound the performance of the TDU and understand for what scenarios and applications it could be used for.

In order to present a complete TDU in this paper, certain design choices (like the choice of motor, battery cells) had to be fixed. For example, ungeared U8 Lite L KV95 direct drive motors controlled with MD80 brushless motor controller drives (MAB Robotics). For applications with higher torque requirements, one could instead spec the AK60-6 V1.1 CubeMars motor from MAB Robotics (3Nm rated / 9Nm peak torque). This would only involve re-sizing the case, modifying the mounting points, and potentially designing a pulley with a different radius if necessary. Designers can in principle use any motor / controller combination that communicates with a CANBUS, with the addition of a CAN transceiver between the Raspberry Pi and the motor controllers, so one would not be limited to motors and controllers from MAB Robotics as well.  

\subsection{Open-Loop Static Torque Performance}

\begin{figure}
    \centering
    \includegraphics[trim={0cm 0cm 0cm 0cm},clip,width=0.95\linewidth]{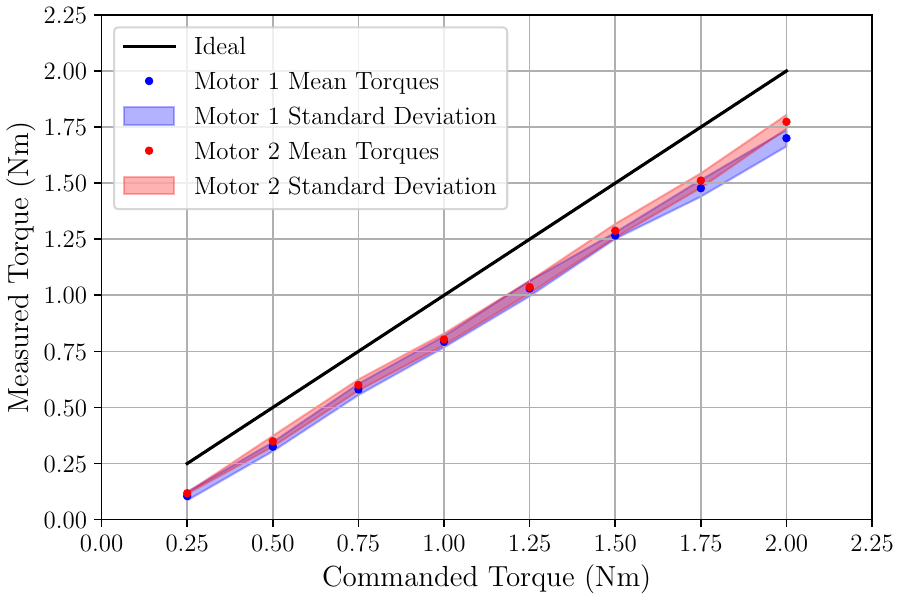}
    \caption{Static torque test results demonstrating the relationship between commanded motor torque and measured output torque based on cable tension measurements.}
    \label{fig:static_torque_test_results}
\end{figure}

 The static torques delivered by the drive are consistent and repeatable, albeit under-deliver consistently (Fig. \ref{fig:static_torque_test_results}). There was a high goodness of fit for both motors, with coefficients of determination of 0.99984 and 0.99975 for the left and right motors, respectively. These correspond to the following relationships: 

\begin{equation}
    \tau_{act} = \biggl\{ 
    \begin{matrix}
        0.915 \cdot \tau_{cmd} - 0.120, \text{left motor} \\
        0.938 \cdot \tau_{cmd} - 0.120, \text{right motor} 
    \end{matrix} 
\end{equation}

where $\tau_{act}$ is the measured motor torque (Nm) and $\tau_{cmd}$ is the commanded motor torque (Nm). This indicates that an inverse linear model may be used to correct the commanded torque depending on the desired output rotor torque, especially if a distal load sensor will not be used to measure cable tension. The offset may come from rotor stiction (since the measurements were made at stall) or from modelling errors on the MD80 controllers. 

% The results from the battery life test also demonstrate that the average motor torque to move the weights is higher than the theoretical 

\subsection{Closed-Loop Velocity Control}

\begin{figure}
    \centering
    \includegraphics[trim={0cm 0cm 0cm 0cm},clip,width=0.95\linewidth]{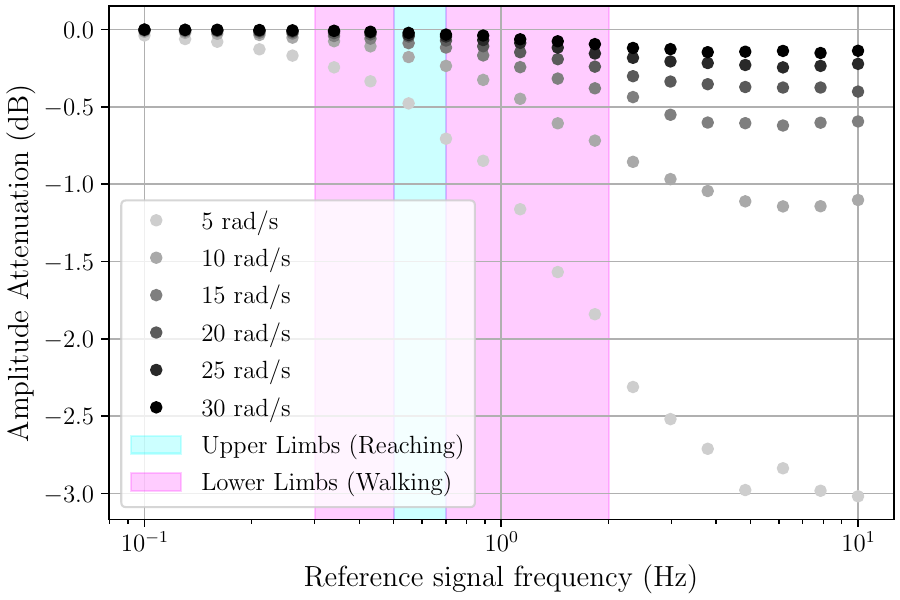}
    \caption{Magnitude Bode plot demonstrating the velocity amplitude attenuation with increasing frequency, across a range of velocity amplitudes.}
    \label{fig:magnitude}

    \centering
    \includegraphics[trim={0cm 0cm 0cm 0cm},clip,width=0.95\linewidth]{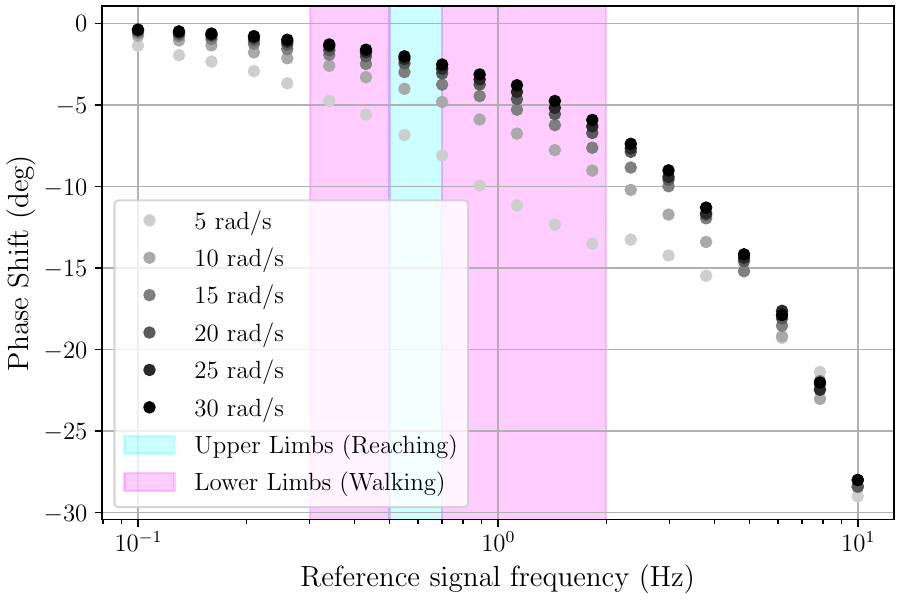}
    \caption{Phase-shift Bode plot demonstrating the increasing lag with increasing driving frequency, across a range of velocity amplitudes.}
    \label{fig:phase}
\end{figure}

For a fixed signal frequency, higher peak rotor speeds had lower attenuation and phase shift (Fig. \ref{fig:magnitude} and \ref{fig:phase}). This is likely a result of using a direct-drive motor and thus the cogging torque acts as a disturbance at lower speeds. For shoulder elevation, Georgarakis and colleagues estimated a minimum movement bandwidth requirement of 0.67Hzm and that elevating the arm to 68$^\circ$ at a peak speed of 171$\frac{^\circ}{s}$ would take 0.4s \cite{georgarakis}. Assuming that an implementation requires 20cm of cable travel (with a 30mm diameter pulley) to elevating the arm in this time, the required rotor speed would be 5.3$\frac{rad}{s}$. At this frequency and speed, the motors perform with almost no signal attenuation, and a phase shift of between 5 and 10 degrees. For lower-limb applications, sources estimate that step frequencies can range between 0.3Hz (incidental stepping) and 2.0Hz (brisk walking) in both healthy and impaired adult populations \cite{tudor,bohannon}. Within this range of frequencies, the phase shift does not exceed 15$^\circ$ across all peak rotor speeds, but for the higher range of walking speeds, amplitude attenuation starts to drop off for the lower range of peak rotor speeds. Ultimately, these results also need to be considered in the context of the final exosuit design, as it depends on the kinematics of the cable routing, supported joints, and the anticipated application.

\subsection{Thermal Cooling}

\begin{figure}
    \centering
    \includegraphics[trim={0cm 0cm 0cm 0cm},clip,width=0.95\linewidth]{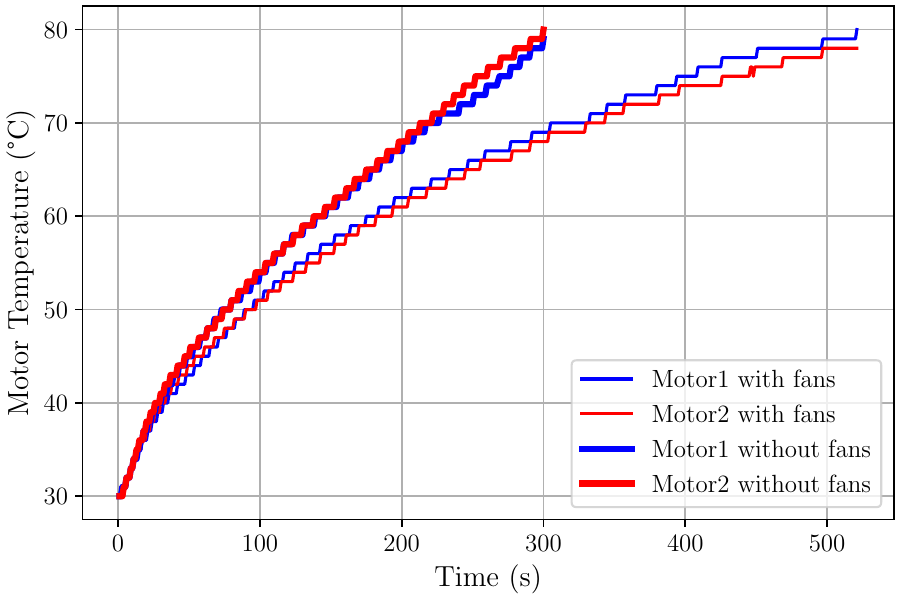}
    \caption{Temperature rise curves for both TDU motors while under a constant dynamic load, with and without the cooling airflow from the side-mounted fans.}
    \label{fig:temperature_rise_results}
\end{figure}

\begin{figure}
    \centering
    \includegraphics[trim={0cm 0cm 0cm 0cm},clip,width=0.95\linewidth]{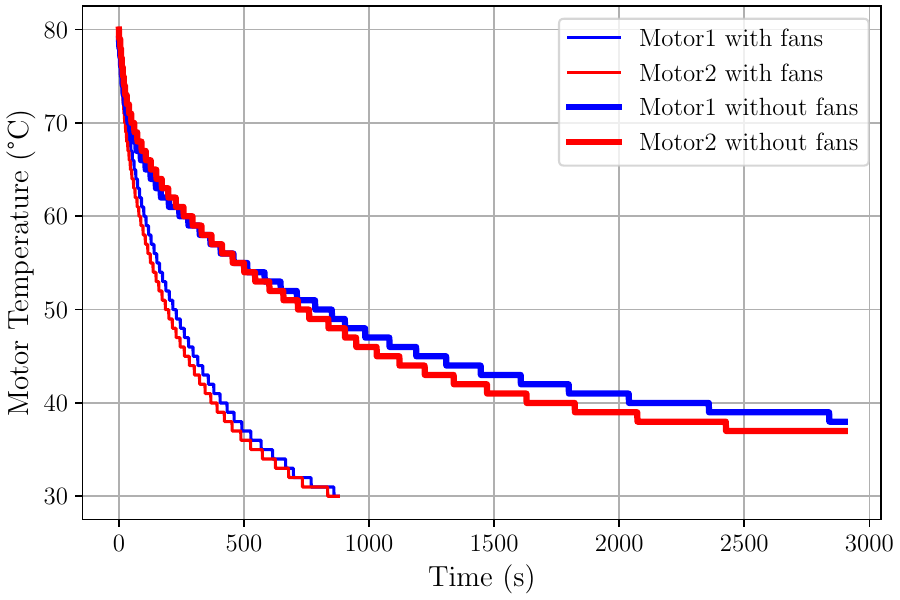}
    \caption{Temperature fall curves for both TDU motors after heated under a constant dynamic load, with and without the cooling airflow from the side-mounted fans.}
    \label{fig:temperature_fall_results}
\end{figure}

The cooling effect of the fans is demonstrated in the temperature rise and fall times (Fig. \ref{fig:temperature_rise_results} and \ref{fig:temperature_fall_results}). During the tests, the ambient air temperature was measured at 23$^\circ$C. Without the fans enabled it took only 300s for the stator temperature to rise from 30$^\circ$C to 80$^\circ$C, while with the fans enabled it took 520s for the same change in temperature. The effect of convective cooling is also apparent during the cooling phase, as the fans could bring the stator temperature from 80$^\circ$C back down to 30$^\circ$C in 850s, whereas with only conductive cooling it took approximately 2000-3000s for the temperatures to fall back down below 40$^\circ$C. These results in absolute terms depend on the ambient air temperature, how much heat is allowed to develop in the stator of the motor, and how fast the fans are controlled. For this particular test, the cutoff was set to 80$^\circ$C, since PLA has a glass transition temperature of approximately 60$^\circ$C (to avoid the case yielding or failing during the test). 
%The motors can be configured to shutdown with stator temperatures as high as 140$^\circ$C. 

Based on the thermal results from this test, it is clear that if the device will be used continuously at torques around 1.5Nm, a combination of active cooling (with fans) and perhaps controlled thermal throttling would be necessary for safe, long-term use (exceeding 10 minutes). This could be done by specifying stronger fans to improve airflow (and thus cooling), but this could come at the cost of increased noise during high-torque performance. Alternatively, one could implement a thermal throttling / rollback of the applied torque when the temperatures exceed a certain threshold (i.e. linearly reduce the output torque from 100\% of commanded torque to 50\% of commanded torque when the temperature is between 80 deg C and 120 deg C respectively). The specific implementation of such a the thermal control law would have to be researched and tested in further detail, but the general idea would be to trade-off performance for thermal management if the system overheats. 

\subsection{Noise}

\begin{figure}
    \centering
    \includegraphics[trim={0cm 0cm 0cm 0cm},clip,width=0.95\linewidth]{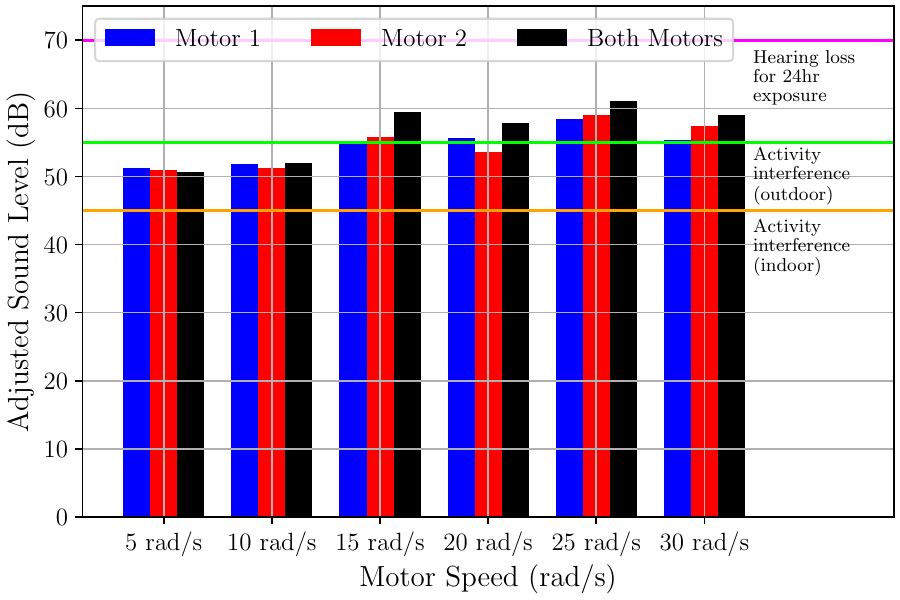}
    \caption{Noise test results for the individual and combined motors across different operating speeds.} 
    \label{fig:noise_test_results}
\end{figure}

With the TDU on and only the fans running, the system had a noise level of 43.7dB. Across all active conditions tested, the TDU demonstrated sound levels ranging from 50.7dB up to 61.1dB. The US EPA identifies a constant 70dB over 24h as the upper level of environmental noise which will not cause any measurable hearing loss over ones life. Furthermore, levels of 55dB outdoors and 45dB indoors are identified as preventing activity interference and annoyance (ie these levels of noise permit spoken conversation and other activities such as sleeping, working and recreation) \cite{EPA}. Thus the current design of this TDU should not over time cause any permanent hearing loss due to use, although during faster movements, the sound produced by the motors might hinder / obstruct conversations. 
%Future designs could leverage thin layers of foam on the slots next to the motors to still allow airflow out of the TDU, but slightly dampen the noise that the motors produce.

\subsection{Battery Life}

\begin{table}[]
    \centering
    \caption{TDU run-time under different loading conditions.}    \begin{tabular}{c|c|c}
        Condition & \makecell{Mean Motor Torque (Nm) \\ $\pm$ Standard Deviation (Nm)} & Runtime (HH:MM:SS) \\
        \hline
        Idle & $8.16\cdot10^{-6} \pm 0.00128$ & 11:20:23 \\ 
        2kg & $0.306 \pm 0.0634$ & 08:06:11 \\ 
        4kg & $0.627 \pm 0.0743$ & 04:39:34 \\
        6kg & $0.943 \pm 0.0941$ & 02:38:20 \\ 
    \end{tabular}
    \label{tab:runtime}
\end{table}

The battery life of the TDU depended greatly on the average external load imposed on it (Table \ref{tab:runtime}), ranging from over 11hr of idle / standby time, down to just over 2.5hr with a constant load of 0.943Nm on both motors. Ultimately, the battery life of the TDU in an exosuit application will also depend on other factors, such as the intended support levels, the user anthropometrics, the transmission efficiency, frequency of use, battery cell chemistry, or if the voltage range is padded. Future designs could also explore custom flat Lipo packs arranged in series and parallel to achieve the desired voltage and capacity. 

\section{Conclusion}
This study presents a series of benchmarking tests that may be used to document the performance capabilities of cable-driven TDUs. The proposed tests capture static torque control, bandwidth of velocity control, thermal management, noise, and battery life. As well, we present a  lightweight and customizable TDU designed for cable-driven wearable applications, showcasing a modular design that allows flexibility in motor selection, pulley configuration, and sensor integration. The modularity of this TDU enables it to be easily adapted for various applications in upper and lower limb exosuits.

Our results reveal that the TDU provides accurate and repeatable torque control. Velocity control has a very high bandwidth, well covering the range of speeds required for upper and lower-limb movements. The cooling system proved effective, with fans extending operational duration under higher loads. Noise measurements indicate that, while the TDU is mostly unobtrusive, improvements could be made to reduce motor sound during high-speed operations. Furthermore, the battery tests confirm adequate runtime for practical, day-long use under moderate loading conditions.

The open sharing of our TDU’s design and testing methodologies contributes to a collaborative approach within the exosuit research community. By making these resources publicly available we seek to promote advancements in assistive wearable technologies.

\section*{Acknowledgments}
The authors would like to thank Michael Herold-Nadig for his continuous support with debugging and integration of the electronics in the TDU. We thank fellow colleagues Alexander Breuss and David Rode for their support with the software and programming, and students Kai Stewart and Matthias Jammot for their contributions to the project. Finally, we acknowledge the constant emotional support of our lab dog Copper. 

\begin{figure}[!htbp]
    \centering
    \includegraphics[trim={0cm 10cm 0cm 10cm},clip,width=0.9\linewidth]{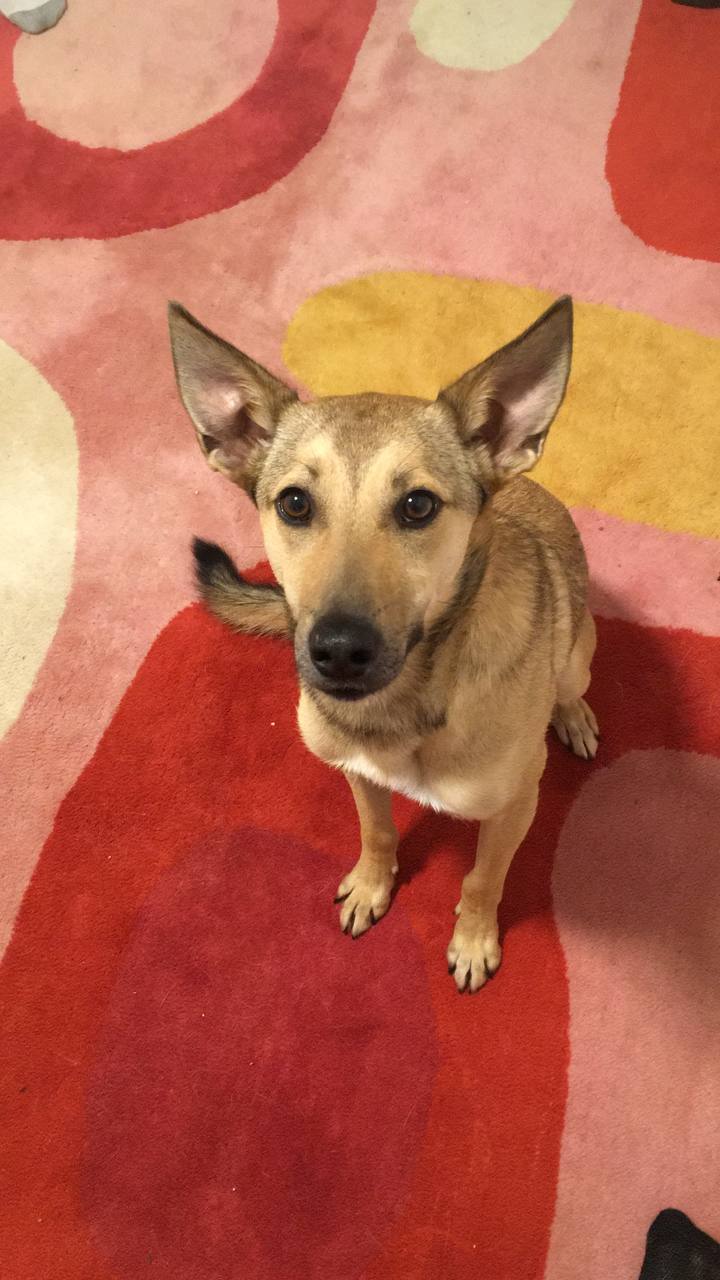}
\end{figure}

\end{document}